\documentstyle[prl,aps]{revtex}
\begin{document}
\def\be{\begin{equation}}
\def\en{\end{equation}}
\def\bq{\begin{eqnarray}}
\def\eq{\end{eqnarray}}
\def\noi{\noindent}
\def\bi{\bigskip}
\def\tr{{\rm tr}}
\def\bc{\begin{center}}
\def\ec{\end{center}}
\def\ii{\'{\i}}

\title{Radiative decays of light vector mesons in a quark level linear sigma 
model}
\author{ M. Napsuciale, S. Rodriguez and E. Alvarado-Anell }
\address {Instituto de F\ii sica, Universidad de Guanajuato,\\
Lomas del Bosque 103, Fracc. Lomas del Campestre \\ 
37150, Le\'on, Guanajuato, M\'exico}

\maketitle

\begin{abstract} 
We calculate the  $P^0 \to \gamma\gamma$, $V^0\to P^0\gamma$
 and $V^0\to {V^\prime}^0\gamma\gamma$ decays in the framework of 
a $U(3)\otimes U(3)$ linear sigma model which includes constituent quarks. 
For the first two decays this approach improves results based on the 
anomalous Wess-Zumino term, with contributions due 
to $SU(3)$ symmetry breaking and vector mixing. The $\phi \to (\omega,\rho)\gamma\gamma$ 
decays are dominated by  resonant $\eta^\prime$ exchange . Our calculation for the later 
decays improves and update similar calculations in the -closely related- 
framework of vector meson dominance. We obtain 
$BR(\phi\to\rho\gamma\gamma)=2.5\times 10^{-5}$ and 
$BR(\phi\to\omega\gamma\gamma)=2.8\times 10^{-6}$  within the scope of the 
high-luminosity $\phi$ factories. 
\end{abstract}

\bi
\bi

\section{Introduction}

Radiative decays of mesons have been a useful tool in the past in  the 
underlying of the dynamics of non-perturbative QCD. In spite of being one 
of the earliest tests for the internal structure of hadrons, 
the interest on this kind of processes is by no means exhausted. 
Indeed, there exists renewed interest on vector meson radiative 
decays \cite{todos} due to the 
possibilities opened by the high luminosity $e^+e^-$ machines which are 
allowing us to study rare decay modes of vector mesons. In particular many 
modes have been recently measured by the experimental groups at Novosibirsk 
\cite{novotodo} and KLOE Coll. at Frascati $\phi$ factory started to release 
even more accurate data \cite{kloe,dafnetodo}. Hopefully this data will help us 
in discriminating the different proposals for the dynamics arising from QCD 
which governs the different channels.  

In the following we study the specific processes $P^0 \to \gamma\gamma$, 
$V^0\to P^0\gamma$ and $V^0\to {V^\prime}^0\gamma\gamma$ decays. In QCD the
$P^0 \to \gamma\gamma$ and $V^0\to P^0\gamma$ decays are induced by the 
contributions of external vector currents to the axial anomalies which can 
be summarized in the Wess-Zumino term \cite{wz}. 
The $V^0\to {V^\prime}^0\gamma\gamma$ decays were 
previously calculated \cite{leibovich} in the framework of heavy vector 
chiral perturbation 
theory (HVCHPT) \cite{hvchpt}. In this theory the 
$V^0\to {V^\prime}^0\gamma\gamma$ decays 
are induced by loops of charged pseudoscalar mesons and the corresponding 
branching ratios turn out to be ratherly small ($\approx 10^-9$).
The importance of intermediate pseudoscalar contributions  to the same 
processes was noticed in \cite{ko}. In this picture  the 
$V^0\to {V^\prime}^0\gamma\gamma$ decays go through the chain 
$V^0\to P^0 \gamma \to {V^\prime}^0\gamma\gamma$ where the anomalous decays 
$V^0\to P^0 \gamma$ play a prominent rol. The most important effect here is 
the  possibility of resonant contributions which 
enhance the corresponding branching ratios and in 
the case of $\phi\to\rho\gamma\gamma$ brings it close to the upper limit 
encountered by the 
CMD-2 Collaboration \cite{novo}. Indeed, the branching ratio calculated in 
\cite{ko} for this decay is 
${\cal B}=1.3\times 10^{-4}$  to be compared with the upper bound reported 
in \cite{novo} ${\cal B}_{exp} < 5\times 10^{-4}$. The possibility of 
improving this bound at DA$\Phi$NE makes worthy to reconsider these decays.

In this work we study the $P^0 \to \gamma\gamma$, 
$V^0\to P^0\gamma$ and $V^0\to {V^\prime}^0\gamma\gamma$ decays
in the framework of
a $U(3)\otimes U(3)$ linear sigma model comprising $U_A(1)$ symmetry breaking 
due to strong dynamics and including constituent quarks. This is a 
generalization to the $U(3)\otimes U(3)$ version of the $U(2)\otimes U(2)$ 
model used in \cite{lucio} to calculate the $\omega\to\rho\pi$ transition and 
the $\omega \to 3\pi$ decay. In this model the contributions to the axial 
anomalies due to external vector fields can be explicitly calculated yielding 
sharp predictions for the $P^0 \to \gamma\gamma$ and $V^0\to P^0\gamma$ 
decays since we work with the physical meson fields. The advantage of the 
explicit introduction of fermion fields is that, on the one hand it makes  
possible to study effects due to $SU(3)$ symmetry breaking and on the other 
hand it also allow us to study the contribution of resonant intermediate 
pseudoscalar states to more involved decays such as 
$V^0\to {V^\prime}^0\gamma\gamma$.

\section{ $U(3)\times U(3)$ Linear Sigma Model with Quarks ($QL\sigma M$)}

The generalization of the model presented in \cite{lucio} to its 
$U(3)\otimes U(3)$ version  can be done by studying the 
$U_{L}(3)\otimes U_{R}(3)$ invariants constructed from $ B=S+ iP$. These 
invariants were studied long ago \cite{olds}
\be
X=tr(BB^{\dagger}),\qquad  
Y=tr(BB^{\dagger})^{2}, \qquad
Z=detB+detB^{\dagger}.
\en
\noi The first two terms are $U_{L}(3)\otimes U_{R}(3)$ invariant whereas the 
$Z$ term  explicitly breaks the $U_{A}(1)$  symmetry but respects 
the $SU_{A}(3) \times U_{V}(3)$ symmetry. The generalization of 
the Lagrangian presented in \cite{lucio} is 
\\
\begin{eqnarray}\label{qlsmlag}
L &=& \bar q[iD_{\mu}\gamma^{\mu}+\sqrt{2}g(S+i\gamma_{5}P)]q + 
\frac{1}{2}tr(D_{\mu}BD^{\mu}B^{\dagger})-
\frac{\mu^{2}}{2}tr(BB^{\dagger})-
\nonumber\\
& & \frac{\lambda}{4}tr(BB^{\dagger})^{2}-
\frac{\lambda^\prime}{4}(tr(BB^{\dagger}))^{2}+tr(CS)-
\beta(detB+detB^{\dagger}).
\end{eqnarray}
\\
\noi where the scalar and  pseudoescalar fields are given by  
$S=\frac{1}{\sqrt{2}}\lambda^{i}s^{i},~~P=\frac{1}{\sqrt{2}}\lambda^{i}p^{i}$,
with $\lambda^{i}$ $i=1...7$ the conventional Gell-Mann matrices and we 
use the flavor matrices $\lambda_{\rm{ns}}=Diag(1,1,0)$ and 
$\lambda_{\rm{s}}=\sqrt{2}Diag(0,0,1)$. Explicitly, the scalar, pseudoscalar 
and vector fields are
\\
\begin{eqnarray}\,
\label{yael}
{S} = \left(
\begin{array}{ccc}
\frac{\sigma_{ns}+a^{0}}{\sqrt{2}} & a^{+} & \kappa^{+} \\
\\
a^{-} & \frac{\sigma_{ns}-a^{0}}{\sqrt{2}} & \kappa^{0} \\
\\
\kappa^{-} & \bar \kappa^{0} & \sigma_{s} \\
\end{array}\right)
\qquad
{P} = \left(
\begin{array}{ccc}
\frac{\eta_{ns}+\pi^{0}}{\sqrt{2}} & \pi^{+} & K^{+} \\
\\
\pi^{-} & \frac{\eta_{ns}-\pi^{0}}{\sqrt{2}} & K^{0} \\
\\
K^{-} & \bar K^{0} & \eta_{s} \\
\end{array}\right)
\qquad
{V_{\mu}} = \left(
\begin{array}{ccc}
\frac{V_{ns}+\rho^{0}}{\sqrt{2}} & \rho^{+} & K^{*+} \\
\\
\rho^{-} & \frac{V_{ns}-\rho^{0}}{\sqrt{2}} & K^{*0} \\
\\
K^{*-} & \bar K^{*0} & V_{s} \\
\end{array}\right)_{\mu}.
\end{eqnarray}
\\

Vector mesons are introduced as external fields through the covariant 
derivatives
\\
\begin{equation}\,
\label{C7}
D_{\mu}q \equiv \left(\partial_{\mu} + 
i\frac{g_{v}}{\sqrt{2}}V_{\mu}\right)q \qquad 
D_{\mu}B\equiv \partial_{\mu}B + i\frac{g_{v}}{\sqrt{2}}[V_{\mu},B].
\end{equation}
\\
The mesonic sector of this theory has been analyzed in detail in 
\cite{masi,tornqvist,andreas}. In particular, the vacuum expectation values 
(vevs) of scalars $a\equiv\frac{<\sigma_{ns}>}{\sqrt{2}}$, 
$b\equiv<\sigma_{s}>$ are related to the pseudoscalar decay constants as
\be
f_{\pi}=\sqrt{2}a \qquad
f_{k}=\frac{1}{\sqrt{2}}(a+b).
\en
As for the fermionic sector, quarks acquire mass 
due to the spontaneous breaking of chiral symmetry. The constituent quark 
mass matrix  generated by this mechanism is $M_{q}=\sqrt{2}g<S>$, where 
$<S>=V=diag(a,a,b)$. Explicitly
\begin{equation}\,
\label{C11}
\begin{array}{ccc}
m_{u}=m_{d}=\sqrt{2}ga=gf_{\pi} \nonumber\\
m_{s}=\sqrt{2}gb=g(2f_{k}-f_{\pi}). \nonumber
\end{array}
\end{equation}
\\
In the calculations of the physical processes to be considered below, the 
free parameters $g,m_{q},m_{s}$ ($q=u,d$) enter in the combinations 
$g/m_{q}$, $g/m_{s}$ which are determined by the weak decay constants of 
pseudoscalars as
\begin{eqnarray}\,
\label{C13}
\frac{g}{m_{q}}=\frac{1}{f_{\pi}} \qquad
\frac{g}{m_{s}}=\frac{1}{2f_{k}-f_{\pi}}.
\end{eqnarray}
\\
The free parameter $g_{v}$ appearing in the covariant derivative (\ref{C7}) 
can be fixed invoking vector meson dominance of the electromagnetic form 
factors of {\it constituent quarks}, i.e. VMD implemented at the constituent 
quark level (QVMD), or directly from the $\rho\to\pi^+\pi^-$ decay. 
QVMD hypothesis was shown to work fairly well in the $U(2)\otimes U(2)$ 
version of the model\cite{lucio}. Under this hypothesis, the isovector 
contribution to the $qq\gamma$ interaction is mediated 
by the rho meson. The corresponding amplitude is
\\
\begin{eqnarray}\,
\label{C14}
{\cal M}_{VMD}=\epsilon^{\alpha}f_{\rho\gamma}\left( \frac{-g^{\alpha\nu}
+q^{\alpha}q^{\nu}/m^{2}_{\rho}}{q^{2}-m^{2}_{\rho}}\right)
\frac{g_{v}}{2}\bar u\gamma_{\nu}u
\end{eqnarray}
\\
where $\epsilon$ stands for the photon polarization vector and 
$f_{\rho \gamma}$ denotes the $\rho -\gamma$ coupling. On the other hand, 
the amplitude for the $qq\gamma$ transition can be characterized on the 
base of Lorentz covariance, gauge invariance, parity etc. as
\\
\begin{eqnarray}\,
\label{C15}
{\cal M}=\epsilon^{\mu}\bar q \left[ e_{q}F_{1}(q^{2})\gamma^{\mu}
+\frac{F_{2}(q^{2})}{m_{q}}q_{\nu}\sigma^{\mu\nu} \right]q
\end{eqnarray}
\\
where $e_{q}$ stands for the quark charge and $F_{1}(q^{2})$, $F_{2}(q^{2})$ 
are form factors which can not be fixed on symmetry arguments alone. Eq. 
(\ref{C14}) has been written such that 
$F_{1}(q^{2})$ is normalized to $F_{1}(0)=1$. In the soft limit 
($q_{\mu}\to 0$) only the charge form factor contributes and comparing with 
Eq.(\ref{C14}) we obtain
\begin{eqnarray}\,
\label{C16}
g_{v}=\frac{em^{2}_{\rho}}{f_{\rho\gamma}}.
\end{eqnarray}
It still remain to fix $f_{\rho \gamma}$. This parameter can be extracted 
from the measured $\rho \rightarrow e^{+}e^{-}$ decay. Using the central 
value for the measured 
branching ratio for this channel \cite{pdg} we obtain $g_v=5.0$. 
As a consistency check for this estimate we can evaluate the isoscalar 
contribution to the $qq\gamma$ interaction mediated by the $\omega$ meson.
A similar analysis for the isoscalar form factor yields
\be
\label{C21}
g_{v}=\frac{em_{\omega}^{2}}{3f_{\omega\gamma}}.
\en
Extracting the $\omega-\gamma$ coupling from the $\omega\to e^+e^-$ decay we 
obtain $g_{v}=5.6$, close to the previous estimate. A calculation for 
the electromagnetic form factor of the $s$ quark yields a 
similar value ($g_v=6.0$). Finally the coupling $g_v$ can also be  
directly extracted from the $\rho^{0}\rightarrow \pi^{+}\pi^{-}$ decay. From 
the Lagrangian (\ref{qlsmlag}) we get $g_{\rho\pi\pi}=g_{v}$. Using the 
measured branching ratio for 
$\rho^{0}\rightarrow \pi^{+}\pi^{-}$ we obtain $g_{v}=6.0 $, fairly 
consistent with the former estimates. In the whole, the vector coupling lies 
in the range $g_v\in [5.0,6.0]$. We will use 
$g_{v}=5.5\pm 0.5$ in the numerical calculations below.

\section{$P\rightarrow \gamma\gamma$ and  $V\rightarrow P\gamma$ decays} 

\subsection{ $P^{0}(p)\rightarrow \gamma (k_1,\varepsilon_1)\gamma 
(k_2,\varepsilon_2) $ decays.}

The $P^{0} \rightarrow \gamma \gamma$ decays are induced by quark loops in 
this model. These contributions are finite and we obtain sharp predictions 
for these decays. The invariant amplitude for the 
$\pi^0\to\gamma\gamma$ decay as calculated in the model is 
\begin{eqnarray}\,
\label{D1}
{\cal M}(\pi^{0}\rightarrow \gamma\gamma)=C_{\pi\gamma\gamma}
\epsilon(\varepsilon_{1},\varepsilon_{2}, k_{1},k_2)~I(p^2 )
\end{eqnarray}
where $\epsilon(p,q,r,l)$ is a shorthand notation for 
$\epsilon_{\mu\nu\rho\sigma} p^{\mu}q^{\nu}r^{\rho}l^{\sigma}$ and 
\begin{eqnarray}\,
\label{D2}
C_{\pi\gamma\gamma} = 8im_{q}gN_{c}(e^{2}_{u}-e^{2}_{d}), 
\end{eqnarray}
\\
with $N_{c}$ the number of colors. The loop integral is given 
by
\\
\begin{eqnarray}\,
\label{D4}
I(p^2 )=\frac{-i}{16\pi^{2}}\int_{0}^{1}dx \int_{0}^{1-x} 
\frac{dy}{m^{2}_{q}\left[1-x(1-x)\frac{p^2}{m^2_{q}} + 
xy \frac{p^2}{m^2_{q}} - i\varepsilon \right]}.
\end{eqnarray}
\\
In the chiral limit ($p^2=m^2_{\pi}=0$) this integral yields
\begin{eqnarray}\,
\label{D5}
I=\frac{-i}{32\pi^{2}m^{2}_{q}}.
\end{eqnarray}
\\
and the invariant amplitude in the chiral limit reduces to 
\be
{\cal M}(\pi^{0}\rightarrow \gamma\gamma) = \frac{\alpha}{\pi f_{\pi}} 
\epsilon(\varepsilon_{1},\varepsilon_{2}, k_{1},k_2)
\en
where we used $\frac{g}{m_{q}}=\frac{1}{f_{\pi}}$. This result coincides with 
the amplitude calculated from the Wess-Zumino term \footnote{This is hardly 
a surprising result since the anomaly  depend on the fermion 
content of the theory. The anomaly matching condition by 't Hooft 
used e.g. in \cite{simon}, is satisfied here due to the same fermionic content 
in the model and in QCD.}. This amplitude yields the decay width
\begin{eqnarray}
\Gamma(\pi^{0}\rightarrow \gamma\gamma)_{QL\sigma M} 
=\frac{\alpha^{2}m^{3}_{\pi}}{64\pi^{3}f^{2}_\pi}=7.63 \ eV
\end{eqnarray}
to be compared with the experimental result 
$\Gamma(\pi^{0}\rightarrow \gamma\gamma)_{exp}=(7.79\pm 0.56) \ eV$.

As for the $\eta\to \gamma\gamma$ and $\eta^\prime\to\gamma\gamma$ we must 
take care of the mixing of the $\eta_{\rm{ns}}$ and $\eta_{\rm{s}}$ in this 
case. The amplitudes  for the physical processes are related to those for the 
flavor fields as follows
\begin{equation}\,
\label{D9} 
\begin{array}{ccc} 
{\cal
M}_{\eta\gamma\gamma}={\cal M}_{\eta_{ns}\gamma\gamma}\cos \phi_{p}- {\cal
M}_{\eta_{s}\gamma\gamma}\sin \phi_{p} \nonumber\\ 
{\cal
M}_{\eta'\gamma\gamma}={\cal M}_{\eta_{ns}\gamma\gamma}\sin \phi_{p}+ {\cal
M}_{\eta_{s}\gamma\gamma}\cos \phi_{p} .\nonumber 
\end{array} 
\end{equation}
\\
The calculation of the amplitude for 
$\eta_{\rm{ns}} \rightarrow \gamma \gamma$ goes along the lines of the case 
of the $\pi^0$. We obtain
\begin{eqnarray}
{\cal M}_{\eta_{ns}\gamma\gamma} = 
C_{\eta_{ns}\gamma\gamma}
\epsilon(\varepsilon_{1},\varepsilon_{2}, k_{1},k_2)
I(m^2_{\eta_{\rm ns}}) 
\end{eqnarray}
where
\begin{eqnarray}
C_{\eta_{ns}\gamma\gamma} &=& 8im_{q}gN_{c}(e_{u}^{2}+e_{d}^{2}).
\end{eqnarray}
In the chiral limit $I$ is given by (\ref{D5}) and the invariant amplitude 
reads
\begin{eqnarray}\,
\label{D7}
{\cal M}(\eta_{ns}\rightarrow \gamma\gamma)=\frac{5\alpha}{3\pi 
f_{\pi}}\epsilon(\varepsilon_{1},\varepsilon_{2}, k_{1},k_2)
\end{eqnarray} 
where we used Eq. (\ref{C13}). Similar calculations for the 
$\eta_{s} \rightarrow \gamma \gamma$ transition yield
\begin{eqnarray}\,
\label{D8}
{\cal M}(\eta_{s}\rightarrow \gamma\gamma)= 
\frac{\sqrt{2}\alpha}{3\pi(2f_{k}-f_{\pi})} 
\epsilon(\varepsilon_{1},\varepsilon_{2}, k_{1},k_2)
\end{eqnarray} 
Finally, the invariant amplitudes for the physical processes 
$\eta \rightarrow \gamma\gamma$ y $\eta' \rightarrow \gamma\gamma$ are
\begin{equation}
\begin{array}{ccc}
{\cal M}_{\eta\gamma\gamma} = 
\frac{\alpha}{3\pi f_{\pi}} \left(5\cos \phi_{p} - 
\frac{\sqrt{2}f_{\pi}}{2f_{k}-f_{\pi}}\sin \phi_{p}\right) 
\epsilon(\varepsilon_{1},\varepsilon_{2}, k_{1},k_2)
\\
{\cal M}_{\eta'\gamma\gamma} = 
\frac{\alpha}{3\pi f_{\pi}} \left(5\sin \phi_{p} +
\frac{\sqrt{2}f_{\pi}}{2f_{k}-f_{\pi}}\cos \phi_{p}\right)
\epsilon(\varepsilon_{1},\varepsilon_{2}, k_{1},k_2)
\end{array} \label{pggamp}
\end{equation}

Notice that these amplitudes reproduce the results of an effective theory 
of mesons in which the contributions of the electromagnetic external sources
to the axial anomalies are introduced using 't Hooft's anomaly matching 
conditions \cite{simon}. As 
discussed in \cite{simon} from Eqs.(\ref{pggamp}) we obtain 
$\phi_{p}\in [38.4^{\circ}, 41.0^{\circ}]$, consistent with other estimates 
\cite{feldmann} and with recent measurements by the KLOE Coll. 
$\phi^{exp}_p=41.8^{+1.9}_{-1.6}$ \cite{kloe}.
 
\subsection{$V(q,\eta)\to P(p)\gamma (k,\varepsilon)$ decays}

The most general form of the invariant amplitude for the 
$V(q,\eta)\rightarrow P(p)\gamma(k,\varepsilon)$ decays is dictated by 
Lorentz covariance, gauge invariance, parity etc. as
\begin{eqnarray}\,
\label{D13}
{\cal M} = A_{vp}
\epsilon(\varepsilon,\eta, k,q)
\end{eqnarray}
where $A_{vp}$ is a form factor. The decay widths, written in terms of 
these form factors are 
\\
\begin{eqnarray}\,
\label{D14}
\Gamma(V\rightarrow P\gamma)=\frac{\vert A_{vp}\vert^{2}m^{3}_{v}}{96\pi}
\left[1-\left(\frac{m_{p}}{m_{v}}\right)^{2}\right]^{3}.
\end{eqnarray}
\\

The value of these form factors on the mass-shell can be extracted from 
the existing measurements for the corresponding decay widths against which 
 the predictions of particular models must be compared. In the $QL\sigma M$ 
the $V^{0}\rightarrow P^{0}\gamma$  decays are induced by quark loops. The 
importance of constituent quark loops in the description of $PVV$ vertices 
was stressed in \cite{mike}. The calculation of the amplitude for the 
$\rho^{0}\rightarrow \pi^{0}\gamma$ decay in the $QL\sigma M$ yields
\\
\begin{eqnarray}
{\cal M} = 
A_{\rho\pi}^{QL\sigma M}\epsilon(\eta,\varepsilon,k,q)
\end{eqnarray}
where
\begin{eqnarray}
A_{\rho\pi}^{QL\sigma M}=4m_{q}g_{v}gN_{c}(e_{u}+e_{d})I(m^2_\rho).
\end{eqnarray}
\\
In this case the loop integral contains spurious imaginary parts coming from 
unphysical on-shell $\bar q q$ channel in the loops. This is due to 
the lack of a confining mechanism for the constituent quarks. We approximate 
this integral to its value at $q^2=0$ and assume that this value is not too 
different from its value at the physical point in the presence of confinement.
This is a reasonable assumption which, as we shall see below, is supported by 
experimental data. Under this assumption the value predicted by the 
QL$\sigma$M for $A_{\rho\pi}$ is
\\
\begin{eqnarray}
A_{\rho\pi}^{QL\sigma M}=\frac{eg_{v}}{8\pi^{2}f_{\pi}}
\end{eqnarray}
\\
In a similar way we calculate the kinematicaly allowed $V\to P^0\gamma$ and 
$P^0\to V^0\gamma$ decays, taking care of the mixing between the 
$V_{\rm{ns}}$ and $V_{\rm{s}}$ to form the physical $\phi$ and $\omega$.
Our results are shown in Table \ref{tableavp} where we defined
\begin{equation} \label{D15}
A=\frac{eg_{v}g}{8\pi^{2}m_{q}}=\frac{eg_{v}}{8\pi^{2}f_{\pi}},~~~~
B=\frac{3eg_{v}g}{8\pi^{2}m_{q}}=\frac{3eg_{v}}{8\pi^{2}f_{\pi}},~~~~
C=\frac{-eg_{v}g}{4\pi^{2}m_{s}}=\frac{-eg_{v}}{4\pi^{2}(2f_{k}-f_{\pi})} 
\end{equation}
We used  $g_{v}=5.5\pm 0.5$ as estimated previously, $f_{\pi}=91.9\pm 3.5
 \ MeV$, $f_k=112.9\pm 1.27 ~ MeV $ \cite{pdg} and
$\phi_{p}=(39.7\pm 1.3)^{\circ}$ in the numerical calculations. The vector 
mixing angle $\phi_{v}$ is extracted from the experimental value for 
$\phi \rightarrow \pi^{0}\gamma$ which yields 
$\phi_{v}=(3.31\pm 0.42)^{\circ}$.
The theoretical uncertainties come mainly from these quantities. The 
experimental amplitudes are obtained from the corresponding branching ratios 
\cite{pdg}.  
\begin{table} 
\begin{center}
\begin{tabular}{llll}
Decay & Amplitude($A_{vp}$)&  $A_{vp}^{QL\sigma M}(GeV^{-1})$ & 
$|A_{vp}^{EXP}|(GeV^{-1})$ \\ \hline
$\rho^{0}\rightarrow \pi^{0}\gamma$ &  $A$ & $~~0.229 \pm 0.022$ & 
$0.294 \pm 0.037$ \\
$\rho^{0}\rightarrow \eta \gamma$ & $Bcos\phi_{p}$ & $~~0.529 \pm 0.053$ &
$0.562 \pm 0.051$ \\
$\omega\rightarrow \pi^{0}\gamma$ & $Bcos\phi_{v}$ & $~~0.687 \pm 0.067$ &
$0.711 \pm 0.016$ \\
$\omega\rightarrow \eta \gamma$ & 
$Acos\phi_{p}cos\phi_{v}+Csin\phi_{p}sin\phi_{v}$& $~~0.164 \pm 0.017$ &
$0.161 \pm 0.013$ \\
$\eta'\rightarrow \rho^{0} \gamma$ & $Bsin\phi_{p}$& $~~0.439 \pm 0.045$ &
$0.388 \pm 0.006$ \\
$\eta'\rightarrow \omega \gamma$ & 
$Asin\phi_{p}cos\phi_{v}-Ccos\phi_{p}sin\phi_{v}$ & $~~0.160 \pm 0.015$ &
$0.137 \pm 0.007$ \\
$\phi\rightarrow \pi^{0}\gamma$* &$Bsin\phi_{v}$& $~~0.039 \pm 0.001$ &
$0.039 \pm 0.001$ \\
$\phi\rightarrow \eta \gamma$ &
$Asin\phi_{v}cos\phi_{p}-Ccos\phi_{v}sin\phi_{p}$ &  $~~ 0.211 \pm 0.020$ &
$0.209 \pm 0.002$ \\
$\phi\rightarrow \eta' \gamma$ &
$Asin\phi_{v}sin\phi_{p}+Ccos\phi_{v}cos\phi_{p}$ & $-0.233 \pm 0.023$ &
$0.225 \pm 0.025$ \\
\end{tabular}
\end{center}
\caption{{\footnotesize Amplitudes for the $V\rightarrow P\gamma$ and  
$P\rightarrow V\gamma$ decays. The process marked as (*) is used as input 
to fix the vector mixing angle.}}
\label{tableavp}
\end{table}

\section{$V\rightarrow V'\gamma\gamma$ decays}

The $V\to V'\gamma\gamma$ decays were firstly studied in \cite{leibovich} in the 
context of Heavy Vector Chiral Perturbation Theory (HVCHPT)\cite{hvchpt}. 
There are two mechanisms contributing to these decays in this formalism. The first
one is the  decay chain $V\to V' P\to V'\gamma\gamma$. The second possibility is 
through loops of charged mesons. 
Furthermore, it was shown in \cite{leibovich} that these mechanisms do not 
interfere in the spin averaged decay rates.

On the experimental side, the decay $\phi\to\rho\gamma\gamma $ has been already tested 
by the CMD-2 Coll. as a byproduct of their analysis of the 
$\phi\to\pi^+\pi^-\pi^0$ decay \cite{novo}. 
Although they are not able to discriminate if the later decay proceeds through the 
chain $\phi\to\rho\pi^0\to\pi^+\pi^-\pi^0$ or it is a direct decay, they can 
distinguish events where the two final photons come from the $\pi^0$ and pose an 
upper bound to the branching ratio for $\phi\to\rho\gamma\gamma$ proceeding through 
other mechanisms. They obtain 
\be
BR(\phi \rightarrow \rho\gamma\gamma)_{exp}<5\times 10^{-4}. \label{upperbound}
\en
\noi In the context of HVCHPT, in addition to the 
$\phi\to\rho\pi^0\to\rho\gamma\gamma$ contributions which have been removed in 
Eq.[\ref{upperbound}], this decay can proceed 
only through loops of charged mesons. The corresponding branching ratios were 
calculated in \cite{leibovich} as
\\
\begin{equation}
\label{D16}
\begin{array}{ccc}
BR(\phi \rightarrow \rho\gamma\gamma)_{HVCHPT}=5.8\times 
10^{-9}(\frac{g_{2}}{0.75})^{4} \nonumber\\
BR(\phi \rightarrow \omega\gamma\gamma)_{HVCHPT}=4.2\times 
10^{-9}(\frac{g_{2}}{0.75})^{4} \nonumber
\end{array}
\end{equation}
\\
where the low energy constant $g_{2}$ is related to the $VV'P$ couplings. 
Estimates for this constant using constituent quark model and  
chiral quark model yield $g_{2}\approx 1$ thus predicting extremely small 
branching ratios.
 
The $\phi \rightarrow (\rho,\omega )\gamma\gamma$ decays were reconsidered 
in \cite{ko} in the framework of Vector Meson Dominance. In this framework, 
in addition to the possibilities appearing in HVCHPT, the 
$V\to V'\gamma\gamma$ decay can proceed through the chain $V\to P\gamma\to 
V'\gamma\gamma$. The contributions from intermediate pseudoscalar 
states enhance the corresponding branching ratios to \cite{ko} 
\\
\begin{equation}\,
\label{D17}
\begin{array}{ccc}
BR(\phi \rightarrow \rho\gamma\gamma)_{VMD}\approx 1.3\times 10^{-4} 
\nonumber\\
BR(\phi \rightarrow \omega\gamma\gamma)_{VMD}\approx 1.5\times 10^{-5} 
\nonumber
\end{array},
\end{equation}
\\
several orders of magnitude larger than
those predicted by HVCHPT. The contributions of intermediate pseudoscalars 
do not appear in HVCHPT due to the conservation of heavy vector meson 
number inherent to this theory. 

Let us consider these decays in the framework of QL$\sigma$M. In this 
formalism the $V\to V^\prime\gamma\gamma$ are induced by quark loops. As 
discussed in the previous section, the assumed low 
variation of the form factors  from $q^2=0\to m^2_V$ is supported by the 
experimental data on $V\to P^0\gamma$ and $P^0\to 
V\gamma$ decays. 
Here we will work with the same assumptions. Under these circumstances, 
our calculation is very similar to that performed in \cite{ko}. However, 
there are some differences which turn out to be relevant in the numerics. 
Firstly, our formalism incorporates  $SU(3)$ symmetry breaking, secondly 
we take into account mixing of vector mesons. The 
$\phi\to (\omega ,\rho )\gamma\gamma$ decays are dominated by the exchange 
of on-shell $\eta^\prime$. Although there are also contributions coming 
from  $\pi$ and $\eta$ exchange, the $\phi\to\pi^0\gamma$ decay is suppressed 
by the OZI rule (see Table I) and 
$\eta$ contributions turn out to be negligible as compared with the on-shell 
$\eta^\prime$ exchange and will not be considered here.

The invariant amplitude for 
$\phi(q_1,\eta_1) \to \rho (q_2,\eta_2) \gamma (k_1,\varepsilon_1) 
\gamma (k_2,\varepsilon_2)$ is 
\begin{eqnarray}\,
\label{D20}
{\cal M}=f_{t}\epsilon(\eta_{1},\varepsilon_{1},k_{1},q_{1})
\epsilon(\eta_{2},\varepsilon_{2},k_{2},q_{2})
+f_{u}\epsilon(\eta_{1},\varepsilon_{2},k_{2},q_{1})
\epsilon(\eta_{2},\varepsilon_{1},k_{1},q_{2})
\end{eqnarray}
where 
\begin{equation}\,
\label{D21}
f_{t}=\frac{A_{\phi\eta'}A_{\rho\eta'}}
{t-\tilde m^{2}_{\eta'}} \qquad
f_{u}=\frac{A_{\phi\eta'}A_{\rho\eta'}}
{u-\tilde m^{2}_{\eta'}} 
\end{equation}
with
\begin{equation}
\tilde m^{2}_{\eta'}= m^{2}_{\eta'}-i~\Gamma_{\eta'} m_{\eta'},~~
s=(q_1-q_2)^2,~~t=(q_{1}-k_{1})^{2},~~u=(q_{1}-k_{2})^{2}. 
\end{equation}
This invariant amplitude corresponds to $\eta'$ exchange in 
$t$ and $u$ channels.  The average squared amplitude is
\be\label{D22}
\vert \bar {\cal M}\vert^{2}=\frac{1}{12}\left[ \vert f_{t}\vert^{2}(t-M^{2})^{2}
(t-m^{2})^{2} + \vert f_{u}\vert^{2}(u-M^{2})^{2}(u-m^{2})^{2}
+Re(f_{t}f^{*}_{u})F(s,t,u)\right]
\en
where
\be
F(s,t,u)=s^2M^2m^2+(M^2m^2-tu)^2
\en
\noi which is explicitly invariant under $t\leftrightarrow u$ as required by 
Bose symmetry. Here,  $M,m $ stand for the masses of $\phi$ and the final vector meson 
respectively. The decay width is given by 

\begin{eqnarray}\,
\label{D24}
\frac{d\Gamma}{ds}=\frac{1}{(2\pi)^{3}}\frac{1}{32M^{3}}\frac{1}{2}
\int_{t_{0}}^{t_{1}}\vert \bar {\cal M}\vert^{2}dt
\end{eqnarray}
where
\begin{equation}
\begin{array}{ccc}
t_{0,1}= \frac{1}{2}[(M^{2}+m^{2}-s)\mp 
\sqrt{(M^{2}+m^{2}-s)^{2}-4m^{2}M^{2}}]. 
\end{array}
\end{equation}
If we integrate numerically this equation using the experimental 
values for the  $\phi\to\eta^\prime\gamma$, 
$\eta^\prime\to\rho\gamma$ and $\eta^\prime\to\omega\gamma$ couplings listed in Table \ref{tableavp}, and 
the values for the masses quoted in  \cite{pdg} we obtain the 
following branching ratios
\be
BR(\phi \rightarrow \rho\gamma\gamma)= 2.3 \times 10^{-5}, \qquad 
BR(\phi \rightarrow \omega\gamma\gamma)= 2.6 \times 10^{-6}. 
\en
\noi If instead we use the values for the couplings as predicted by the 
model also listed in Table \ref{tableavp} we get 
\be
BR(\phi \rightarrow \rho\gamma\gamma)_{QL\sigma M}=2.5 \times 10^{-5}, \qquad 
BR(\phi \rightarrow \omega\gamma\gamma)_{QL\sigma M}=2.8 \times 10^{-6}. 
\nonumber
\end{equation}

These results are roughly five times smaller than those reported in \cite{ko}. 
It is instructive to understand where the differences come from. To this end 
let us analyze the $\phi\eta^\prime\gamma$ coupling in both schemes. 
In our formalism, this coupling is given by
\be
g_{\phi\eta^\prime\gamma}= \frac{eg_{v}}{8\pi^{2}f_{\pi}}sin\phi_{v}sin\phi_{p}-
\frac{eg_{v}}{4\pi^{2}(2f_{k}-f_{\pi})}cos\phi_{v}cos\phi_{p},
\en
whereas the coupling used in \cite{ko}, when we use $\phi_p=\theta +\alpha$ where 
$sin\alpha=\sqrt{2/3}$, $cos\alpha=\sqrt{1/3}$, can be written as
\be
g_{\phi\eta^\prime\gamma}= \frac{eg_{v}}{4\pi^{2}f_{\pi}}cos\phi_{p}.
\en
Clearly, modulo an irrelevant global sign, our analytical results reduce to those 
presented in \cite{ko} when we take  the SU(3) limit ($f_k=f_\pi$) and consider 
ideal mixing for vector mesons $\phi_v=0$. However, even if we take $\phi_v=0$, 
the SU(3) corrections introduce a factor 
$(\frac{2f_k-f_\pi}{f_\pi})^2=(1.44)^2= 2.04$ in the branching ratios.  
Further sources of discrepancy due to SU(3)symmetry breaking enter in the value 
used in the numerics for the pseudoscalar mixing angle. The extraction of this 
angle from $\eta\to\gamma\gamma$ and $\eta^\prime \to\gamma\gamma$ is also 
sensitive to SU(3) symmetry breaking as can be seen in Eq.[\ref{pggamp}]. Our 
result, $\phi_p = 39.7\pm 1.3$, is consistent with the recent measurement by 
the KLOE Coll. $\phi^{exp}_p = 41.8^{+1.9}_{-1.6}$ \cite{kloe} and is larger 
than the one used in \cite{ko}, namely  $\phi_p = \theta +54.7^\circ = 34.7^\circ $. 
Additional sources of discrepancy come from  the non-ideal vector mixing angle 
and the numerical value used for $g_v$. The importance of all these effects is 
clearly exhibited in the calculation of the branching ratio for the 
$\phi\to\eta^\prime \gamma$ decay.
Indeed, this branching ratio was calculated to be $2.1 \times 
10^{-4}$ in \cite{ko}. In our formalism we obtain $6.8 \times 10^{-5}$ to be 
compared with the recent measurement by the KLOE Coll. \cite{kloe}, 
namely $BR(\phi\to\eta^\prime \gamma)_{exp}=(6.1\pm 0.61\pm 0.43)\times 10^{-5}$. 
In summary, SU(3) corrections are necessary for the correct description of 
$\phi\eta^\prime\gamma $, hence they are also relevant for 
$\phi\to (\rho,\omega )\gamma\gamma$ which are dominated by on-shell 
$\eta^{\prime}$ exchange. For the latter decays additional sources for the 
different numerics come from the $\eta^{\prime}\rho\gamma$ and 
$\eta^{\prime}\omega\gamma$ couplings, whose ratio deviates from the value of 3 
predicted by SU(3)(see Table I), and from phase space integral.

In any case, the $\phi \to \rho\gamma\gamma$ decay is still 
close to the upper bound in Eq.(\ref{upperbound}) and both decays are 
within the reach of DA$\Phi$NE.

\section{Conclusions}

In this work we study the decays $P^0\to\gamma\gamma$, $V^0\to P^0\gamma$
 and $V^0\to {V^\prime}^0\gamma\gamma$, in the framework of 
an $U(3)\otimes U(3)$ linear sigma model which includes constituent quarks.
All these decays are induced by quark loops in this model. 
For the $P^0\to\gamma\gamma$ decays we reproduce results previously obtained 
considering the contributions of external electromagnetic fields to the 
axial anomalies in the framework of the  $U(3)\otimes U(3)$ linear 
sigma model and using 't Hooft's anomaly matching conditions \cite{simon}. In 
the case of $V^0\to P^0\gamma$ decays there are spurious contributions coming 
from the opening of on-shell $\bar q q$ pairs in the loops  due to the 
lack of a mechanism for confinement. We avoid this problem assuming that the 
corresponding form factors are smoothly varying functions of $q^2$ and 
calculating their value at $q^2=0$. Comparison with experimental results shown 
in Table \ref{tableavp} support this picture. The calculations for  
$V^0\to {V^\prime}^0\gamma\gamma$ under these approximations improve those  
performed in \cite{ko} by incorporating effects  due to $SU(3)$ symmetry 
breaking and the mixing of vector mesons. Our numerical results turn out to be 
smaller than those obtained in \cite{ko} roughly by a factor of 5. 
This difference can be traced to $SU(3)$ symmetry breaking, non-ideal mixing 
for vector mesons and numerical input -which manifests in their large prediction for the 
branching ratio of the $\phi\to\eta^\prime \gamma$ decay- and also to phase space 
numerical integration. Our predictions for branching ratios of the most interesting 
processes are:  
$BR(\phi\to\rho\gamma\gamma)_{QL\sigma M}=2.5\times 10^{-5}$, 
$BR(\phi\to\omega\gamma\gamma)_{QL\sigma M}=2.8\times 10^{-6}$. In the former 
case the calculated $BR$ is quite close to the upper bound obtained 
by the CMD-2 Coll. 
$BR(\phi \rightarrow \rho\gamma\gamma)_{exp}<5\times 10^{-4}$ \cite{novo} and 
these decays could be detected for the first time at DA$\Phi$NE.

\section{Acknowledgments}
Work supported by Conacyt-Mexico under project 37234-E.

\end{document}